\documentclass[%
twocolumn,
superscriptaddress,
 amsmath,amssymb,
 aps,
prc,
floatfix,
]{revtex4-1}

\usepackage{amsmath}
\usepackage{amssymb}
\usepackage{amsfonts}

\usepackage{xcolor}

\usepackage{dsfont}

\usepackage{braket}
\usepackage{graphicx}
\usepackage{dcolumn}
\usepackage{bm}

\begin{document}

\preprint{APS/123-QED}

\title{Gamow Shell Model description of ${ {  }^{ 40 } }$Ca(d,p) transfer reaction} 

\author{A. Mercenne}
\affiliation{Center for Theoretical Physics, Sloane Physics Laboratory, Yale University, New Haven, Connecticut 06520, USA}
\author{N. Michel}
\email{nicolas.michel@impcas.ac.cn}
\affiliation{Institute of Modern Physics, Chinese Academy of Sciences, Lanzhou 730000, China}
\author{J.P. Linares Fern\'{a}ndez}
\affiliation{Grand Acc\'el\'erateur National d'Ions Lourds (GANIL), CEA/DSM - CNRS/IN2P3, BP 55027, F-14076 Caen Cedex, France}
\author{M. P{\l}oszajczak}
\affiliation{Grand Acc\'el\'erateur National d'Ions Lourds (GANIL), CEA/DSM - CNRS/IN2P3, BP 55027, F-14076 Caen Cedex, France}

\date{\today}

\begin{abstract}
  Transfer reactions are essential to determine spectroscopic factors and astrophysical reaction rates. However, their theoretical evaluation is typically effected using standard reaction theory, from which structure degrees of freedom are absent. While reaction cross sections have been implemented in the frame of the no-core shell model with continuum, this model can be applied in practice only to the lightest nuclei.
  The use of the core + valence nucleon picture is then necessary to include inter-nucleon correlations in reaction cross sections involving medium nuclei. For this, we will use the recently developed coupled-channel Gamow Shell Model (GSM-CC) for direct reactions and extend it to the evaluation of transfer cross sections. As an example, we will study the ${ {  }^{ 40 } }$Ca(d,p) transfer reaction with GSM-CC. Experimental data can be successfully reproduced, but at the price of the use of a very phenomenological Hamiltonian.
\end{abstract}

\maketitle


\section{Introduction}
{
 Nuclear reaction rates are one of the most important ingredients in describing many astrophysical phenomena. 
 However, direct measurements of the cross sections at stellar energies are very challenging, especially when it involves charged particles.
 Instead, transfer reactions can be used as an indirect method to access the desired reaction rates of astrophysical interest \cite{bardayan_2016,nunes_2020}. 
 More particularly, (d,p) reactions have been extensively used to extract spectroscopic factors for astrophysically relevant isotopes to constrain neutron-capture rates \cite{thomas_2005,kozub_2005,thomas_2007,kozub_2012,manning_2019}.
 Indeed, one-nucleon transfer reactions are the probe of choice to obtain information about the nuclear response to nucleon addition (single-particle strength) as a function of energy, angular momentum and parity.
 Traditionally, one-nucleon transfer reactions to bound states can be used to extract information regarding direct capture, where those populating the continuum are used to extract the resonant and/or compound capture.

 To describe the thousands of reactions of astrophysical interest at the relevant energies, one has to rely on theoretical approaches.
 The properties of radioactive nuclei, underpinning the nuclear mechanisms involved in astrophysical processes, are strongly affected by couplings to many-body continuum of scattering and decay channels.  
 Therefore, a unified theory of these nuclei involves a comprehensive description of bound states, resonances and scattering many-body states within a single theoretical framework, and this is one of the main goals of the nuclear theory. 
 A pioneer work in this direction was initiated with the continuum shell model \cite{rotter_review_1991,Okolowicz2003,bennaceur_2000,rotureau_2005,rotureau_2006,volya_2005}, and has been extended to \textit{ab initio} description of structure and reactions of light nuclei within the no-core shell model coupled with the resonating-group method (NCSM/RGM) \cite{quaglioni_2008,quaglioni_2009} and the no-core shell model with continuum (NCSMC) \cite{baroni_2013a,baroni_2013b}.
 Deuteron induced transfer reactions to s-shell and p-shell have been investigated with NCSM/RGM \cite{navratil_2012,raimondi_2016} in the context of primordial and stellar nucleosynthesis. 
 In the same effort to unify nuclear structure and reactions, progress have been made in the development of microscopic \textit{ab initio} optical potentials for deuteron induced transfer reactions \cite{rotureau_2020}.

 An alternative approach to describe radioactive nuclei within a unifying framework has been proposed with the open quantum system formulation of the shell model, the Gamow Shell Model (GSM) \cite{michel_review_2009,michel_book_2021}.
 GSM offers the most general treatment of couplings between discrete and scattering states, as it makes use of Slater determinants defined in the Berggren ensemble \cite{berggren_1968} of single-particle states.
 For the description of scattering properties and reactions, it is convenient to formulate GSM in the representation of reaction channels (GSM-CC) \cite{jaganathen_2014}. 
 The Hamiltonian of GSM-CC is Hermitian because matrix elements are calculated in the harmonic oscillator basis. However, the calculation of resonances using this Hamiltonian is done in the Berggren basis, so that the Hamiltonian matrix in GSM-CC becomes complex symmetric. The cross sections are calculated by coupling the real-energy incoming partial waves to the target states given by the Hermitian Hamiltonian. Consequently, the framework related to cross section calculation is fully Hermitian, whereas complex energies arise for resonances because one diagonalizes the complex symmetric Hamiltonian matrix induced by Berggren basis representation. 
 GSM in the coupled-channel representation, which is based on the RGM, has been applied to the description of ${ {  }^{ 6 } }$Li via deuteron induced elastic scattering on ${ {  }^{ 4 } }$He \cite{mercenne_2019}.

 To benchmark GSM-CC for transfer reactions, we apply it to (d, p) reactions on the doubly magic stable $^{40}$Ca nuclei.
The Letter is organized as follows. The general formalism of the GSM-CC is briefly introduced in Chap.~\ref{Formalism}. The Hamiltonian and results of GSM-CC calculations are presented in Chap. \ref{discussions}. In particular, the low-energy spectra and binding energies of $^{41}$Ca, $^{41}$Sc, $^{42}$Ca, $^{42}$Sc, $^{42}$Ti  are discussed and the description of elastic cross sections: $^{40}$Ca(n,n)$^{40}$Ca,  $^{40}$Ca(p,p)$^{40}$Ca, and transfer cross section $^{40}$Ca(d,p)$^{41}$Ca is presented. Finally, conclusions are summarized in Chap. \ref{conclusions}.
}

\section{Theoretical framework}
\label{Formalism}
In this section, we will briefly outline the GSM-CC formalism, as more details can be found in  Refs.~\cite{jaganathen_2014,mercenne_2019,michel_book_2021}.

We work in the cluster orbital shell model (COSM) formalism \cite{PhysRevC.38.410}, i.e.~that all space coordinates are defined with respect to that of a given inert core:
\begin{equation}
\mathbf{r} = \mathbf{r_{\rm lab}} - \mathbf{R_{\rm CM}^{\rm (core)}} \label{COSM_coordinate}
\end{equation}
where $\mathbf{r_{\rm lab}}$ is the space coordinate in the laboratory frame, $\mathbf{R_{\rm CM}^{\rm (core)}}$ is the center of mass coordinate of the inert core in the laboratory frame, and $\mathbf{r}$ is the COSM space coordinate. More precisely, $\mathbf{r_{\rm lab}}$ and $\mathbf{r}$ are respectively the laboratory and COSM valence nucleon coordinates in the case of one-nucleon systems, while they correspond to the center of mass coordinate in the laboratory and COSM frames, respectively, in the case of a many-nucleon cluster projectile. The fundamental advantage of COSM is that it is translationally invariant, as COSM space coordinates are clearly relative, so that no spurious center of mass excitation can occur therein \cite{PhysRevC.38.410,michel_book_2021}.

The ${ A }$-body state of the system is decomposed into reaction channels :
\begin{equation}
  \ket{ { \Psi }_{ M_{ A}  }^{ J_{ A}  } } = \sum_{ c } \int_{ 0 }^{ +\infty } \ket{{ \left( c , r \right) }_{ M_{ A}  }^{ J_{ A}  } } \frac{ { u }_{ c }^{J_{ A} M_{ A} \textbf{}} (r) }{ r } { r }^{ 2 } ~ dr \; ,
  \label{scat_A_body_compound}
\end{equation}
where the radial amplitude ${ {u}_{ c }^{J_{ A} M_{ A} }(r) }$, describing the relative motion of the projectile with respect to the core in a channel ${ c }$, is the solution to be determined for a given total angular momentum ${J_{ A} }$ and its projection ${M_{ A} }$. Note that, in case of a cluster channel, the coordinates of the cluster nucleons are not present in Eq. (\ref{scat_A_body_compound}), as $r$ is the COSM radial coordinate of the cluster center of mass. This is in contrast with usual formulations of channels in other models and embodies the cluster approximation used in GSM-CC. In fact, we do not need to explicitly treat nucleonic coordinates inside the composite projectile, as we restrict our GSM-CC basis to the two-cluster mass partitioning case. However, it is possible to effectively include deuteron break-up by including channels that involve intrinsic scattering states of the deuteron calculated using the Berggren basis. This was done in Ref.\cite{mercenne_2019} where the scattering reaction $^4$He(d,d) is considered.

The channel states are defined as:
\begin{equation}
  \ket{ \left( c , r \right)}  = \hat{ \mathcal{A}} \ket{ \{ \ket{\Psi_{\rm T }^{J_{\rm T }} } \otimes \ket{r ~ \ell ~ J_{\text{int}} ~ J_{\rm P}} \}_{ M_{ A}  }^{J_{ A} }} \ . \\ 
  \label{channel}
\end{equation} 
The channel index $c$ stands for the partitions and quantum numbers $\{ A - a , { J }_{\rm T } ; a , { L } , J_{\text{int}}, J_{\rm P}\}$, and ${\hat{ \mathcal{A}}}$ is the inter-cluster antisymmetrizer that acts among the nucleons pertaining to different clusters. 
The states $\ket{\Psi_{\rm T }^{J_{\rm T }} }$ and $\ket{r ~ \ell ~ J_{\text{int}} ~ J_{\rm P}}$ are the target and projectile channel states with their associated total angular momentum ${ { J }_{\rm T } }$ and ${ { J }_{ \rm P } }$, respectively.
The angular momentum couplings read $\mathbf{J}_{ P } = \mathbf{J}_{\rm int } + \pmb{\ell}$ and  ${ \mathbf{J}_{ A} = \mathbf{J}_{\rm P} + \mathbf{J}_{\rm T}}$.

The coupled-channel equations can then be formally derived from the Schr{\"o}dinger equation: $H \ket{\Psi_{M_{ A }}^{J_{ A }}} = E \ket{\Psi_{M_{ A }}^{J_{ A }}}$, as:
\begin{equation}
  \sum_{\rm c}\int_{0}^{\infty}  \!\!\! r^{ 2 } \left( H_{ \rm cc' } (r , r') - E N_{\rm cc' } (r , r') \right) \frac{ { u }_{\rm c } (r) }{ r } = 0	\ ,
  \label{cc_cluster_eq}
\end{equation}
with ${ E }$ the scattering energy of the ${ A }$-body system, and where the kernels are defined as:
\begin{align}
  & H_{\rm cc' } (r,r') = \bra{ ({\rm c},r) } \hat{ H } \ket{({\rm c'},r') } \label{h_cc_compound} \\
  & N_{\rm cc' } (r,r') = \braket{ ({\rm c},r) | ({\rm c'},r') } \label{n_cc_compound}
\end{align}
For sake of clarity, we have dropped the total angular momentum labels ${ J_{ A}  }$ and ${ M_{ A}  }$, but one should keep in mind that the resolution of Eq. (\ref{cc_cluster_eq}) is done for fixed values of ${J_{ A} }$ and ${ M_{ A}  }$.

Due to the decoupling of the target and projectile at high energy, it is more convenient to express the Hamiltonian $\hat{ H }$ as simply:
\begin{equation}
  \hat{ H } = \hat{ H }_{\rm T } + \hat{ H }_{ \rm P } + \hat{ H }_{\rm TP }	\ ,	
  \label{new_hamiltonian}
\end{equation}
where ${ \hat{ H }_{\rm T } }$ and ${ \hat{ H }_{ \rm P } }$ are the Hamiltonians of the target and projectile, respectively, while the inter-cluster Hamiltonian ${ { \hat{ H } }_{\rm TP } }$ is defined as: ${ { \hat{ H } }_{\rm TP } = \hat{ H } - { \hat{ H } }_{\rm T } - { \hat{ H } }_{ \rm P } }$, where ${ \hat{ H } }$ is considered here as a standard shell model Hamiltonian.

To be more specific, ${ { \hat{ H } }_{\rm T } }$ is the center-of-mass free intrinsic Hamiltonian of the target, and its eigenvectors are ${ \ket{ { \Psi }_{\rm T }^{ { J }_{\rm T } } } }$ with eigenvalues ${ { E }_{\rm T }^{ { J }_{\rm T } } }$.
The projectile Hamiltonian is then given by ${ { \hat{ H } }_{ \rm P } }$, and can be decomposed as follow: ${ { \hat{ H } }_{ \rm P } = { \hat{ H } }_{ \text{int} } + { \hat{ H } }_{ \text{CM} } }$, where ${ { \hat{ H } }_{\rm int } }$ describes its intrinsic properties and ${ { \hat{ H } }_{\rm CM } }$ the movement of its center-of-mass, defined in a single channel $c$, and where one radial coordinate $r$ occurs:

\begin{equation}
   { \hat{ H } }_{\rm CM } = \frac{ { \hbar }^{ 2 }}{ 2 \Tilde{m}_{\rm P}} \left( -\frac{ { d }^{ 2 }}{ d r^{ 2 }} + \frac{\ell(\ell + 1)}{r^2} \right)  + { U_{\rm CM}^{ \ell }} (r) \ ,
  \label{HCM_definition}
\end{equation} 
where ${ {\Tilde{m} }_{ \rm P } }$ in this equation is the reduced mass of the projectile and ${ { U }_{\rm CM }^{ \ell }(r) }$ is the basis-generating Woods-Saxon (WS) potential for nucleon projectile, while it is the weighted sum of proton and neutron basis-generating WS potentials for deuteron wave functions \cite{mercenne_2019,michel_book_2021}. Its central and spin-orbit parts $U^{ \ell }_{{\rm CM}, {\rm C}} (r)$ and $U^{ \ell }_{\rm CM, SO } (r)$ read in the latter case:
\begin{eqnarray}
\!\!\!\!\!\!\!\!\!\!\!\! U^{ \ell }_{{\rm CM}, {\rm C}} (r) &=& U^{ \ell }_{\rm p, C} (r) + U^{ \ell }_{\rm n, C} (r)  
\label{UCM_central} \\
\!\!\!\!\!\!\!\!\!\!\!\! U^{ \ell }_{\rm CM, SO } (r) &=& \frac{1}{2}~U^{ \ell }_{\rm p, SO} (r) + \frac{1}{2}~U^{ \ell }_{\rm n, SO} (r) 
 \label{UCM_so_average} \ ,
\end{eqnarray}
where $U^{ \ell }_{\rm p, C} (r)$, $U^{ \ell }_{\rm p, SO} (r)$ and $U^{ \ell }_{\rm n, C} (r)$, $U^{ \ell }_{\rm n, SO} (r)$ are the WS basis-generating central and spin-orbit potentials for proton and neutron, respectively.
The potential $U^{ \ell }_{\rm CM}(r)$ of Eq. (\ref{hamiltonian_matrix_elmts}) then reads:
\begin{eqnarray}
U^{ \ell }_{\rm CM} (r) &=& U^{ \ell }_{\rm CM, C} (r) \nonumber \\
&+& \frac{1}{2} ~ { U^{ \ell }_{\rm CM, SO }} (r) ~ (\pmb{\ell} \cdot \mathbf{J_{\rm int}}) \ ,
\label{UCM}
\end{eqnarray}
where an additional 1/2 factor in the spin-orbit part arises because the deuteron nucleons have an orbital angular momentum approximately equal to $\ell/2$ \cite{mercenne_2019,michel_book_2021}.

In order to calculate the kernels Eqs. (\ref{h_cc_compound}) and (\ref{n_cc_compound}), one expands ${ \ket{ (c,r) } }$ onto a one-body Berggren basis:

\begin{equation}
  \ket{ (c,r) } = \sum_{n} \frac{ { u }_{ n \ell} (r) }{ r } \ket{ (c,n) }    \ ,
  \label{expansion_channel_n}
\end{equation}
where ${ \ket{ ({\rm c},n) } = \hat{ \mathcal{A}} \ket{ \{ \ket{\Psi_{\rm T }^{J_{\rm T }} } \otimes \ket{n ~ \ell ~ J_{\text{int}} ~ J_{\rm P}} \}_{ M_{ A}  }^{J_{ A} }} }$, with ${ { \hat{ H } }_{\rm CM } \ket{ n \ell } = { E }_{\rm CM } \ket{ n \ell } }$, implying that ${ n }$ refers to the projectile center-of-mass shell number in the Berggren basis state. Spin-dependence has not been added in the $ { u }_{ n \ell} (r)$ wave function notation for simplicity. The basis of ${ \ket{ n \ell } }$ states is then generated by diagonalizing ${ { \hat{ H } }_{\rm CM } }$.
Note that ${ \ket{ { J }_{ \text{int} } } }$ is the eigenvector of ${ { \hat{ H } }_{ \text{int} } }$ with the eigenvalue ${ { E }_{\rm P}^{ { J }_{ \text{int} } } }$.

Consequently, we can expand Eq. (\ref{h_cc_compound}) onto the basis of ${ \ket{ (c,n) } }$ using Eq. (\ref{expansion_channel_n}) and derive the following expression for the Hamiltonian kernel:

\begin{eqnarray}
  H_{\rm cc' } (r, r') &=& \left( { \hat{ H } }_{\rm CM } + E_{\rm T }^{ { J }_{\rm T } } + {E}_{\rm P}^{ { J }_{ \text{int} } } \right) \frac{ \delta (r - r') }{ r r' } { \delta }_{\rm cc' } \nonumber \\
  &+&  { \tilde{ V }}_{\rm cc' } (r , r')
  \label{hamiltonian_matrix_elmts}
\end{eqnarray}
where  $\tilde{ V }_{\rm cc' }(r,r')$ includes the remaining short-range potential terms of the Hamiltonian kernels and is the inter-cluster potential. Note that $H_{\rm cc' } (r, r')$ reduces to its diagonal part at large distance as $\tilde{ V }_{\rm cc' }(r,r')$ vanishes identically at a large but finite radius outside the target. Hence, nucleon transfer, which is induced by $\tilde{ V }_{\rm cc' }(r,r')$, and consequently ${ { \hat{ H } }_{\rm TP } }$, can only occur in the vicinity of the target and not in the asymptotic region.

Clearly, the determination of ${ { \tilde{ V } }_{\rm cc' }(r,r') }$ involves the calculation of the matrix elements of ${ { \hat{ H } }_{\rm TP } }$, which contain a shell model Hamiltonian.
In order to compute ${ { \hat{ H } }_{\rm TP } }$, one has to expand each ${ \ket{ ({\rm c},n) } }$ onto a basis of Slater determinants built upon single-particle (s.p.) states of the Berggren ensemble.
In practice, the intrinsic target and projectile states, ${ \ket{ { \Psi }_{\rm T }^{ { J }_{\rm T } } } }$ and ${ \ket{ { J }_{\rm int } } }$ respectively, are already calculated with that basis, as ${ { \hat{ H } }_{\rm T } }$ and ${ { \hat{ H } }_{ \text{int} } }$ are solved using the GSM.
Note that, in general, as we deal with very light projectiles, ${ { \hat{ H } }_{ \text{int} } }$ is solved within a no-core framework, and this will be the case in the present study.
The remaining task consists in expanding ${ \ket{ n ~ \ell ~ { J }_{\rm int } } }$ in a basis of Slater determinants. In GSM-CC, this is done by applying a center-of-mass excitation raising operator onto ${ \ket{ { J }_{\rm int } } }$. 
More details can be found in \cite{mercenne_2019,michel_book_2021}.

The many-body matrix elements of the norm kernel Eq. (\ref{n_cc_compound}) are calculated using the Slater determinant expansion of the cluster wave functions ${ \ket{ ({\rm c},n) } }$.
The treatment of the non-orthogonality of channels is the same as in the one-nucleon projectile case \cite{jaganathen_2014}.
Note that the antisymmetry of channels, enforced by the antisymmetrizer in Eq. (\ref{channel}), is exactly taken into account through the expansion of many-body targets and projectiles with Slater determinants.

Once the kernels are computed, the coupled-channel equations (\ref{cc_cluster_eq}) can be solved using a numerical method based on a Berggren basis expansion of the Green's function ${ { (H - E) }^{ -1 } }$, that takes advantage of GSM complex energies. Details of this method can be found in Refs. \cite{mercenne_2019,michel_book_2021}.

One remarks that a formulation of coupled-channel equations based on Berggren basis expansions has been formulated in Ref.  \cite{IDBETAN201418} and applied therein to the calculation of the deuteron bound state and phase shifts.

\section{Results and discussions}
\label{discussions}
We consider a $^{40}$Ca core with one or two valence nucleons to study the $^{40}$Ca(d,p) reaction. All partial waves up to $\ell=4$ are included in the model space of GSM and GSM-CC. As all considered target nuclei are well bound, it is sufficient to define model spaces with HO wave functions in GSM, i.e.~GSM reduces to standard shell model for target structure. For this, one includes all HO wave functions bearing $n \leq 5$ in $s, p, d, f, g$ partial waves above the $^{40}$Ca core. The HO length used is 1.88 fm. No truncation is imposed. As described in Sec. \ref{Formalism}, the many-body resonant and scattering wave functions in GSM-CC are expanded in a Berggren basis of reaction channels.

The core of the Hamiltonian in both GSM and GSM-CC is mimicked by a WS potential and the residual interaction between nucleons is the Furutani-Horiuchi-Tamagaki (FHT) interaction \cite{1979Furutani}. Basis-generating potentials in GSM-CC are also of WS type. 

All WS potentials possess a diffuseness $d = 0.65$ fm and a radius $R_0 = 1.27 A^{1/3} = 4.34$ fm, 
so that they differ only by the central and spin-orbit potential depths, denoted respectively by $V_{\rm o}$ and $V_{\rm so}$. 

The single-particle states of $^{41}$Ca and $^{41}$Sc, that is the $7/2^-_1$, $3/2^-_1$, $5/2^-_1$ and $1/2^-_1$ states, correspond to the one-body states of the $fp$ shell. Thus, the core WS potentials for partial waves $\ell=1,3$ have been fitted to reproduce the single-particle states of $^{41}$Sc and $^{41}$Ca, respectively. Core potentials for $\ell=0,2,4$ partial waves play a very small role in the structure of $A=40-42$ nuclei, hence they cannot be determined on experimental energies. Therefore, their values have been fitted to reproduce the reaction cross sections. WS core potential depths are listed in Tab. \ref{core_WS_parameters}.

\begin{table}[htb] 
\centering
\caption{The central ($V_{\rm o}$) and spin-orbit ($V_{\rm so}$) potential depths of the core WS potentials in proton $(\rm p)$ and neutron parts $(\rm n)$ for partial waves $\ell=0,\dots,4$  (in MeV). \label{core_WS_parameters} }
\begin{tabular}{l|cccccc|}\hline  \hline
Depth & $\ell = 0$  & $\ell = 1$  & $\ell = 2$  & $\ell = 3$  & $\ell = 4$  \\ \hline 
$V_{\rm o}(\rm p)$    &  50  & 55.142  & 62     & 56.601  & 56.5  \\ \hline 
$V_{\rm so}(\rm p)$ & ---  & 6.012   & 5      & 2.884   & 7  \\ \hline 
$V_{\rm o}(\rm n)$    & 60.5 & 54.302  & 59.5   & 54.204  & 40  \\ \hline 
$V_{\rm so}(\rm n)$ & ---  & 5.007   & 2      & 2.850   & 2  \\ \hline  \hline 
\end{tabular}
\end{table}

\begin{figure*}
     \vskip -4truecm
    \includegraphics[width=2.2\columnwidth]{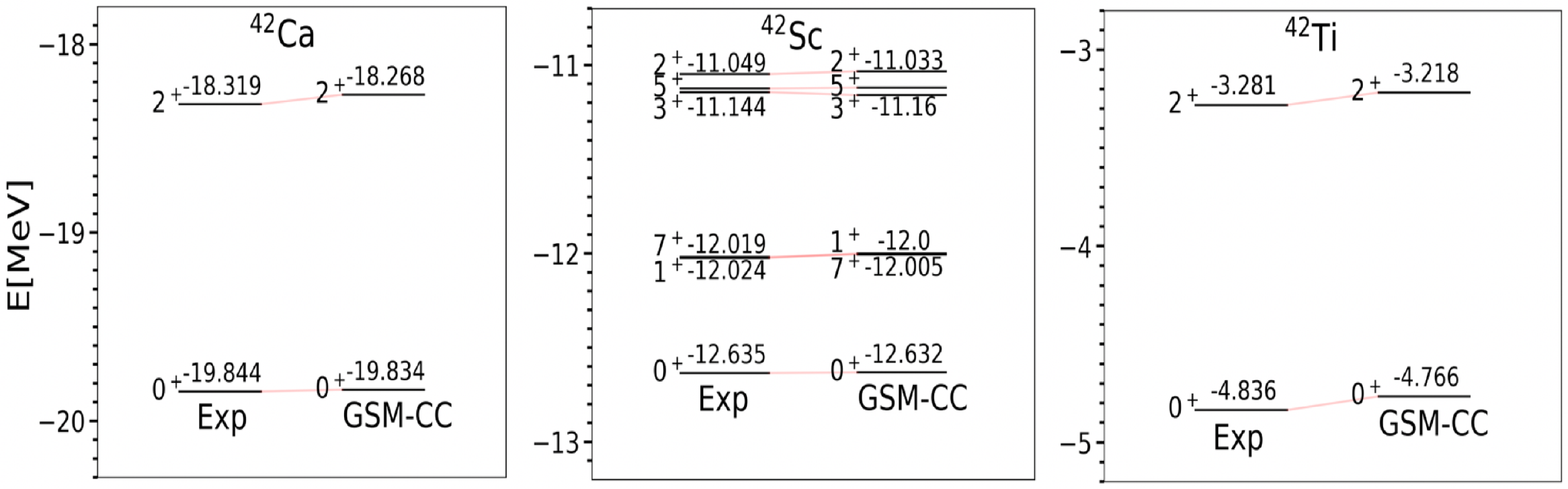}
    \vskip -4.8truecm
    \caption{The low-energy positive-parity states of $^{42}$Ca,  $^{42}$Sc and $^{42}$Ti nuclei calculated using the  GSM-CC are compared to the experimental data \cite{ensdf}.}
    \label{42_spectra}
\end{figure*}
The parameters of the FHT interaction have been fitted to reproduce the low-lying spectra of $A=42$ nuclei in GSM and GSM-CC and are listed in Tab. \ref{FHT}. 
\begin{table}[htb] 
\centering
\caption{\label{Table.InterParamFHT} The optimized parameters of the FHT interaction consist of central ($V_{\rm c}^{ST}$), spin-orbit, ($V_{\rm LS}^{ST}$) and tensor ($V_{\rm T}^{ST}$) coupling constants \cite{jaganathen_2017}.
$S=0,1$ and $T=0,1$ are the spin and isospin of the two nucleons, respectively.
Parameters are given in MeV for the central and spin-orbit parts, and in MeV fm$^{-2}$ for the tensor part. \label{FHT} }
\begin{tabular}{cccccccc}\hline \hline 
$V_{\rm c}^{11}$  & $V_{\rm c}^{10}$ &$V_{\rm c}^{00}$ & $V_{\rm c}^{01}$ & $V_{\rm LS}^{10}$  & $V_{\rm LS}^{11}$  & $V_{\rm T}^{10}$  & $V_{\rm T}^{11}$ \\ \hline 
17.745 &  $-$4.516  & $-$0.210 &  $-$7.386 & $-$2509 & 1000 & $-$4.102 & $-$0.257 \\ \hline \hline
\end{tabular}
\end{table}
\begin{table}[htb] 
\centering
\caption{Same as Tab. \ref{core_WS_parameters}, but for the basis-generating WS potentials \label{basis_WS_parameters} }
\begin{tabular}{l|cccccc}\hline \hline 
Depth & $\ell = 0$  & $\ell = 1$  & $\ell = 2$  & $\ell = 3$  & $\ell = 4$  \\ \hline 
$V_{\rm o}(\rm p)$    & 61   & 60     & 43     & 56      & 59  \\ \hline 
$V_{\rm so}(\rm p)$ & ---  & 6.681  & 5      & 2.854   & 5  \\ \hline 
$V_{\rm o}(\rm n)$    & 65   & 54.302 & 65.13  & 54.204  & 65  \\ \hline 
$V_{\rm so}(\rm n)$ & ---  & 5.007  & 3.9    & 2.850   & 5  \\ \hline  \hline
\end{tabular}
\end{table}

\begin{table}[htb] 
\centering
\caption{The comparison of GSM-CC separation energies ${\rm S}_{\rm n}$, ${\rm S}_{\rm p}$,
${\rm S}_{\rm 2n}$, ${\rm S}_{\rm 2p}$, and 
${\rm S}_{\rm d}$ with experimental ones \cite{ensdf}. Energies are given in units of MeV.
\label{sep_energies} }
\begin{tabular}{l|ccccccc}\hline \hline 
Nucleus & ${\rm S}_{\rm n}^{\rm (th)}$  & ${\rm S}_{\rm n}^{\rm (exp)}$ & ${\rm S}_{\rm 2n}^{\rm (th)}$ & ${\rm S}_{\rm 2n}^{\rm (exp)}$ & ${\rm S}_{\rm d}^{\rm (th)}$ & ${\rm S}_{\rm d}^{\rm (exp)}$
\\ \hline 
$^{41}{\rm Ca}$    &  8.38  & 8.363 & - & - & - & - \\ \hline 
$^{42}{\rm Ca}$ & 11.36  & 11.48 & 19.74 & 19.8 & - & - \\ \hline 
$^{42}{\rm Sc}$    & 11.42 & 11.55 & -  & - & 12.53 & 12.48 \\ 
  \hline \hline
Nucleus & ${\rm S}_{\rm p}^{\rm (th)}$  & ${\rm S}_{\rm p}^{\rm (exp)}$ & ${\rm S}_{\rm 2p}^{\rm (th)}$ & ${\rm S}_{\rm 2p}^{\rm (exp)}$ & ${\rm S}_{\rm d}^{\rm (th)}$ & ${\rm S}_{\rm d}^{\rm (exp)}$ \\ \hline   
$^{41}{\rm Sc}$     & 1.11  & 1.09 & - & - & - & - \\
\hline 
$^{42}{\rm Ti}$ & 3.57  & 3.75 & 4.68 & 4.83 & - & - \\
\hline 
$^{42}{\rm Sc}$    & 4.15 & 4.27 & -  & - & 12.53 & 12.48 \\ \hline \hline
\end{tabular}
\end{table}
The deuteron projectile is issued from a GSM calculation using a Berggren basis defined with two-body relative coordinates, whereby the N$^3$LO interaction is diagonalized (see also Ref. \cite{michel_book_2021} for calculations of diproton, dineutron and deuteron observables in that framework). One has to generate antisymmetric composite states by adding deuteron wave functions to target states afterwards. However, as the latter are defined in HO model spaces with laboratory coordinates, deuteron wave functions must be expanded in the same basis.
For this, the bound and scattering deuteron eigenstates issued from Berggren basis diagonalization are firstly expanded in a basis of HO states defined with two-body relative coordinates and then in a basis of HO states in laboratory coordinates using Talmi-Brody-Moshinsky coefficients. For the latter operation, one uses an HO basis defined in a 10 $\hbar \omega$ space. The transformation from laboratory coordinates to COSM coordinates is neglected, as it can be shown that its effect is minimal compared to the other theoretical approximations present in our model \cite{michel_book_2021}. This allows to recapture the overall structure of deuteron eigenstates. The energy of the deuteron ground state is fixed at its experimental energy of -2.2 MeV, which is close to its theoretical value, equal to -2.1 MeV. Deuteron break-up can be taken into account by including the scattering deuteron eigenstates arising from the Berggren basis diagonalization, as was done in Ref.\cite{mercenne_2019}. However, as we only consider a small deuteron projectile energy in the following, of about 1.8 MeV, along with an inert $^{40}$Ca target core, deuteron break-up cannot occur in our present calculations due to energy conservation, so that only the deuteron ground state will be included in deuteron channels.

As only nucleon projectiles are present in $^{40}$Ca(p,p) and $^{40}$Ca(n,n) reactions, the GSM-CC reaction channels $[^{40}{\rm Ca}(0^+_1)\otimes{\rm p}(L_j)]^{J^{\pi}}$, 
$[^{40}{\rm Ca}(0^+_1)\otimes{\rm n}(L_j)]^{J^{\pi}}$
are directly defined and solved in the GSM-CC Berggren basis. However, this is not possible when considering the $^{40}$Ca(d,p) reaction, where one has to use the HO basis to build all composite states, as we saw for $^{40}$Ca+d channels. Thus, in this case, proton and neutron wave functions are also expanded with the HO basis so as to form the composite basis HO states of $^{41}$Ca + p and $^{41}$Sc + n.

The HO composite basis states are used only for the generation of the potentials entering Eq. (\ref{hamiltonian_matrix_elmts}), where HO energy truncation is effected at $E_{\rm max}^{(\rm HO)} = 8$ $\hbar \omega$. The GSM-CC Hamiltonian of Eq. (\ref{hamiltonian_matrix_elmts}) is solved afterwards using the GSM-CC Berggren basis (see Sec.~II D of Ref.\cite{mercenne_2019}). The GSM-CC Berggren basis of proton, neutron and deuteron projectiles is generated by WS potentials whose parameters are listed in Tab. \ref{basis_WS_parameters} (see Sec. \ref{Formalism} for formulas). 

The Berggren basis of protons and neutrons consists of all partial waves up to $\ell=4$. Those for deuteron consist of ${^3S_1}, {^3P_0}, {^3P_1}, {^3P_2}, {^3D_1}, {^3D_2}, {^3D_3}$ channels.
Consequently, the composite channels are those of $^{40}$Ca$(0_1^+)$ + d, $^{41}$Ca($K^{\pi})$ + p and $^{41}$Sc($K^{\pi})$ + n, where $K^{\pi}=1/2^-, 3/2^-, 5/2^-$ and $7/2^-$. 

For the calculation of cross sections, the $J^\pi$ quantum numbers of composites are restricted to $0^-$, $1^-$, $2^-$, $1^+$, $2^+$ and $3^+$. Contours are discretized with 21 Gauss-Legendre points for protons and neutrons and 30 Gauss-Legendre points for deuterons. Contours are defined with $k_{\rm peak} = 0.2-0.01i$, $k_{\rm middle} = 0.4-0.01i$ and $k_{\rm max} = 2$ fm$^{-1}$. Corrective factors have been added in positive-parity channels, which are equal to 1.14, 0.99, 1.08, 1.04, 1.1 and 1.12 in $0^+$, $1^+$, $2^+$, $3^+$, $5^+$ and $7^+$ channels, respectively. They allow to reproduce the low-energy, positive-parity spectra of $^{42}$Ca, $^{42}$Sc, $^{42}$Ti nuclei (see Fig.(\ref{42_spectra})).
Note that their influence on the $^{40}$Ca(d,p) cross section is minimal.
The larger values of the corrective factors for $0^+$, $5^+$, and $7^+$ channels might be due to the absence of $0^+_2$, $3^-_1$, and $2^+_1$ low-energy core excitation in $^{40}$Ca.




\begin{figure}
    \centering
    \includegraphics[width=1.0\columnwidth]{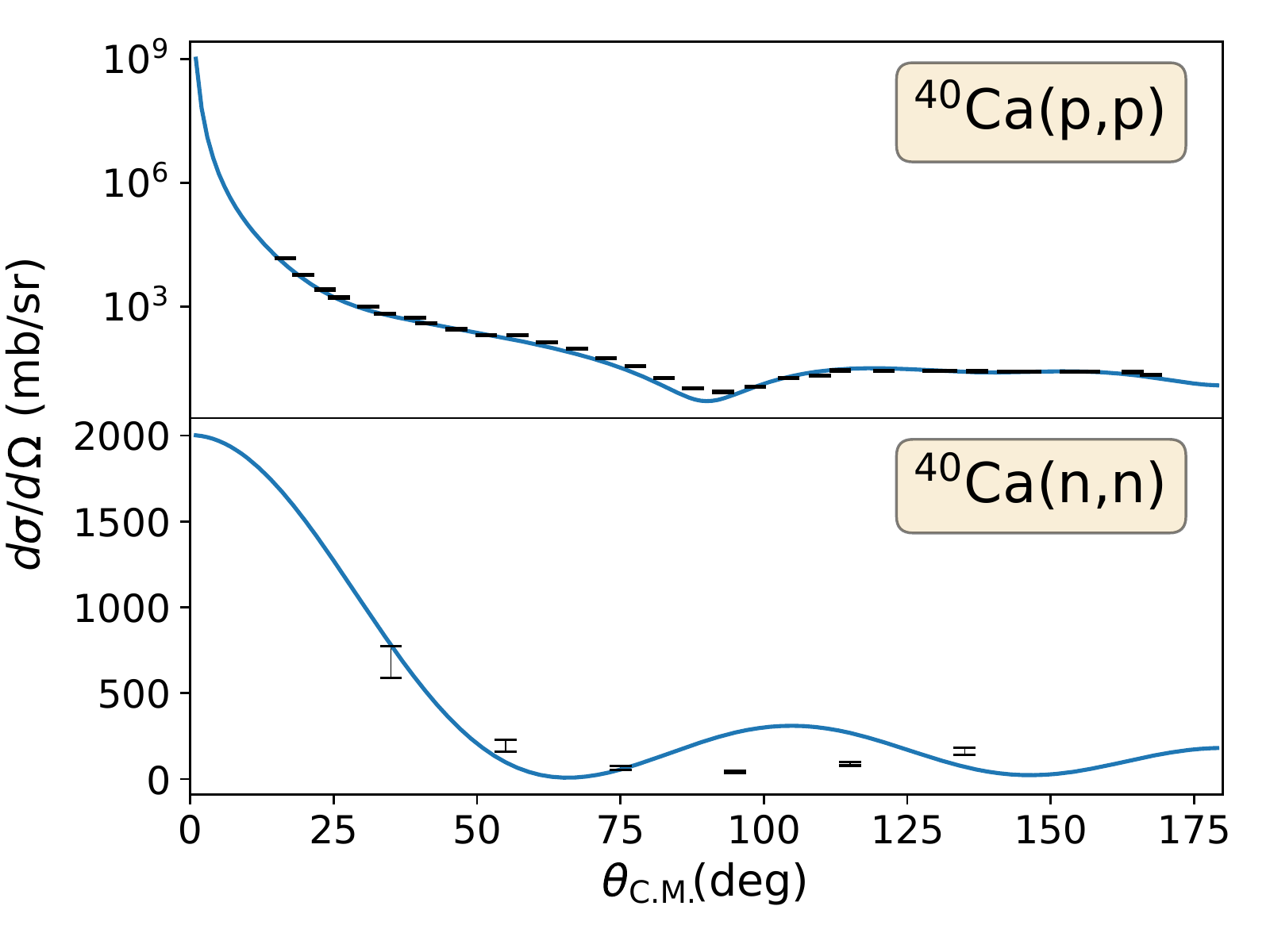}
    \caption{Cross sections $^{40}$Ca(p,p) at CM energy 9.61 MeV and $^{40}$Ca(n,n) at CM energy 2.69 MeV. Cross section and angle are given in CM coordinates. Experimental data for cross sections $^{40}$Ca(p,p) and $^{40}$Ca(n,n) are taken from Refs.\cite{PhysRevC.4.1130,Winterhalter_1971}, respectively. }
    \label{40Capp-nn}
\end{figure}

Results for one- and two-nucleon separation energies ${\rm S}_{\rm n}$, ${\rm S}_{\rm p}$,
${\rm S}_{\rm 2n}$, ${\rm S}_{\rm 2p}$, and ${\rm S}_{\rm d}$ calculated in GSM-CC for $A=41$ and 42 nuclei are compared with experimental separation energies in Tab. \ref{sep_energies}. 
The excited states of $A=42$ nuclei are reproduced in GSM-CC with a typical precision of $\sim$50 keV. However, their impact on considered cross sections is minimal.

\begin{figure}
    \centering
    \includegraphics[width=1.0\columnwidth]{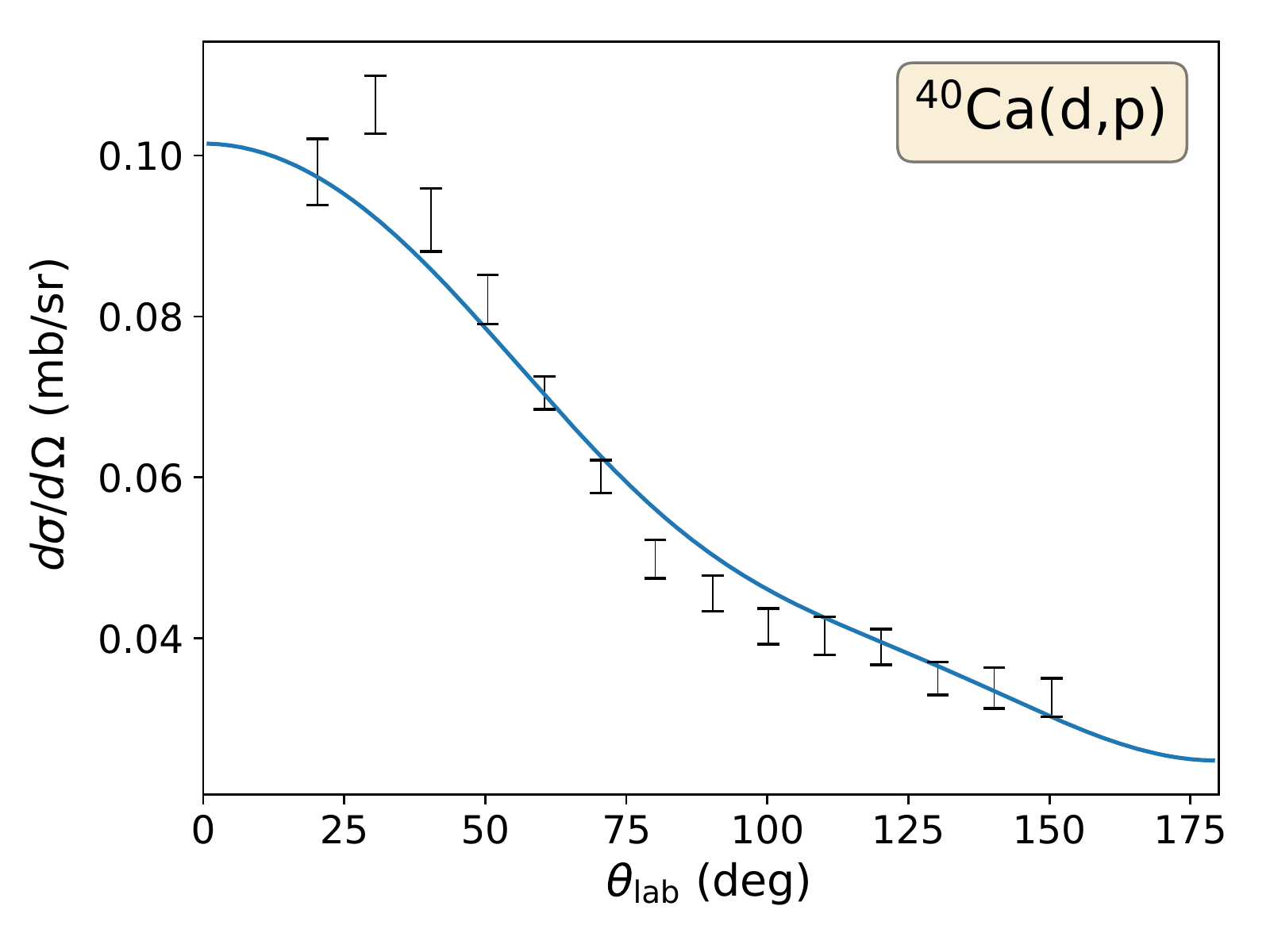}
    \caption{Cross section $^{40}$Ca(d,p) at CM energy 1.853 MeV. Besides projectile energy, cross section and angle are given in laboratory coordinates. Experimental data  are taken from Ref.\cite{Fodor_1965}.}
    \label{40Cadp}
\end{figure}

The calculated cross sections are compared with the data in Figs. \ref{40Capp-nn} and \ref{40Cadp}. 
The GSM-CC transfer cross section $^{40}$Ca(d,p) is very well reproduced both in the form and the magnitude.
Also the calculated proton and neutron elastic cross sections reproduce experimental data very well, except for $^{40}$Ca(n,n) at large angles $\theta_{\rm CM}$ where the GSM-CC cross section exhibits small oscillations.
It is possible to determine a WS neutron core potential reproducing correctly the $^{40}$Ca(n,n) cross section in the full range of $\theta_{\rm CM}$. However, in this case, the $^{40}$Ca(d,p) cross section cannot be reproduced satisfactorily, so that we preferred not to optimally fit the $^{40}$Ca(n,n) cross section for other cross sections to be well reproduced. 

Note that it is possible to obtain a good reproduction of the $^{40}$Ca(d,p) cross section data only by fine tuning both the Hamiltonian parameters and the WS basis-generating potential describing the continuum wave functions and their asymptotes, which is essential for a correct description of the cross sections. Indeed, $^{40}$Ca(d,p) cross sections vary by large factors in an energy interval centered on 1.853 MeV of a few hundreds of keV \cite{brown_1970}. Thus, important cancellations occur between all partial waves. Hence, in order to compensate for these intricate phenomena, the parameters of WS basis-generating potential had to be fitted as well to reproduce the experimental data.

\section{Conclusions}
\label{conclusions}
The microscopic description of reaction cross sections demands to use models where structure and reaction degrees of freedom are present. This is the case in GSM-CC, where target and projectile wave functions are calculated with shell model. Scattering wave functions can be evaluated afterwards from coupled-channel equations, defined with the microscopically calculated reaction potentials. However, contrary to direct reactions, transfer reactions are very rarely studied at microscopic level. 

Thus, we studied the $^{40}$Ca(d,p) transfer reaction cross section in GSM-CC, along with associated direct nucleon scattering reactions $^{40}$Ca(p,p) and $^{40}$Ca(n,n). This is, up to our knowledge, the first calculation of this type in heavier nuclei combining shell model and coupled-channel equation approaches. We could obtain a good overall reproduction of experimental cross sections, except for $^{40}$Ca(n,n) at large angle. The $^{40}$Ca(d,p) transfer reaction cross section is particularly well described. However, this came at the price of having to very precisely fine tune Hamiltonian and Berggren basis parameters. 

Consequently, while the first GSM-CC calculation applied to transfer reaction cross section is satisfactory from a phenomenological point of view, it still remains very difficult in practice to systematically study transfer reaction cross sections with the GSM-CC. The renormalization of neglected reaction channels using complex coupling channel-channel
potentials seems to be necessary in the future for that matter.

\begin{acknowledgments}
 This work has been supported by the National Natural Science Foundation of China under Grant Nos. 12175281 and 11975282; the Strategic Priority Research Program of Chinese Academy of Sciences under Grant No. XDB34000000; the State Key Laboratory of Nuclear Physics and Technology, Peking University under Grant No. NPT2020KFY13.

\end{acknowledgments}

\bibliographystyle{apsrev4-1}
\bibliography{refs}

\end{document}